\documentclass[a4paper,12pt]{article}
\usepackage{graphicx}
\usepackage[english]{babel}
\usepackage{amssymb,amsbsy,amsmath,eucal}
\usepackage[compress]{cite}
\newcommand{\ha}{\mbox{\small$\frac{1}{2}$}}
\newcommand{\qu}{\mbox{\small$\frac{1}{4}$}}
\newcommand{\im}{\,{\rm i}\,}

\newcommand{\lab}[1]{\label{#1}}
\newcommand{\re}[1]{(\ref{#1})}
\newcommand{\nn}{\nonumber}
\newcommand{\B}[1]{\boldsymbol{#1}}
\newcommand{\s}[1]{\mathsf{#1}}

\newcommand{\inta}{\hspace*{-.1cm}\int \hspace*{-.08cm}}
\newcommand{\intab}{\hspace*{-.2em}\int \hspace*{-.4em}\int \hspace{-.2em}}
\newcommand{\D}[2]{{\rm d}^{#1}{#2}\,}
\begin{document}

\title{Solvable two-particle systems with time-asymmetric
interactions in de Sitter space}

\author{A. Duviryak}
\date{Institute for Condensed Matter Physics of NAS of Ukraine,\\
1 Svientsitskii Street, Lviv, UA-79011, Ukraine \\
              Tel.: +380 322 701496, \
              Fax: +380 322 761158\\
              {duviryak@icmp.lviv.ua}           
}

\maketitle

\begin{abstract}
The two-particle models in de Sitter space-time with time-asymmetric
retarded-advanced interactions are constructed. Particular cases of
the field-type electromagnetic and scalar interactions are
considered. The manifestly covariant descriptions of the models
within the Lagrangian and Hamiltonian formalisms with constraints
are proposed. It is shown that the models are de Sitter-invariant
and integrable. An explicit solution of equations of motion is
derived in quadratures by means of projection operator technique.
\end{abstract}


\section{Introduction}
\renewcommand{\theequation}{1.\arabic{equation}}
\setcounter{equation}{0}

It is known that one has to deal with complex
difference-differential equations when considering a relativistic
classical dynamics of a system of interacting charges
\cite{Roh90,YaT12}. This is even more the case for scalar
\cite{YaT12}, gravitational \cite{BDDIM81} or non-Abelian
\cite{D-R81} interactions where the dynamics is governed by
integro-differential equations. Such a hereditary dynamics is
neither solvable nor appropriate for the Hamiltonian description. In
order to avoid these difficulties, Staruszkiewicz \cite{Sta70}, Rudd
and Hill \cite{R-H70} invented the model describing the following
time-asymmetric interaction of two pointlike charged particles: the
advanced field of the first particle acts on the second particle,
the retarded field of the second particle acts on the first
particle, and a radiation reaction is neglected. This model is built
of the action-at-a-distance Tetrode-Fokker variational functional
\cite{Tet22,Fok29} via replacing its integrand, the symmetric Green
function of d'Alembert equation, by the retarded (or advanced) one.
In this way the model was reformulated to the Lagrangian form and
then to the Hamiltonian form \cite{Sta71} which was shown integrable
\cite{Kun74} due to exact Poincar\'e-invariance. The
Staruszkiewicz-Rudd-Hill model was generalized for a variety of
non-electromagnetic time-asymmetric interactions (scalar,
gravitational, confining etc.) \cite{Ste85,Duv97,Duv98}, and
corresponding quantum versions \cite{D-S01,Duv01} revealed their
physical adequacy, despite of an artificially broken causality of
interactions.

A dynamics of interacting particles in a curved space-time is
more complicated than that in the Minkowski space. To author's knowledge,
solvable examples of two-particle dynamics are unknown even for the cases
of symmetric space-times which occur in cosmology.
However, the  Staruszkiewicz-Rudd-Hill
model can serve as a basis for appropriate generalization.
Here a class of two-particle systems with time-asymmetric
interactions is considered in de Sitter space-time. It includes, in
particular, the models with electromagnetic \cite{D-Y19E} and scalar interactions
built in terms of appropriate Green functions. The representation of
de Sitter space-time as a hyperboloid in the 5-dimensional Minkowski
space ${\Bbb M}_5$ is exploit. A single particle dynamics is used to
introduce elements of this representation. Then for a dynamical
system of two particles with the electromagnetic, scalar and more general
time-asymmetric interactions the covariant variational principle is constructed, and an
appropriate Hamiltonian description with constraints is developed.
The dynamics is invariant with respect to de Sitter group O(1,4)
and integrable. A solution for this dynamics is obtained
in terms of quadratures. This is done by means of projection
operators built in terms of conserved canonical generators of
O(1,4). The system of free particles as a time-asymmetric model is
particularly considered.


\section{Manifestly covariant test particle mechanics in de Sitter space}
\renewcommand{\theequation}{2.\arabic{equation}}
\setcounter{equation}{0}

Let us start with the action integral determining the dynamics of a
test particle of the mass $m$ in a curved space-time:

%
\begin{equation}\lab{2.1}
I=
-m\int\limits_{\tau_1}^{\tau_2}\!\!\D{}{\tau}\sqrt{g_{\mu\nu}(x(\tau))\dot{x}^\mu(\tau)\dot{x}^\nu(\tau)};
\end{equation}
here $\tau$ parameterizes points $x(\tau)$ of a particle world line, i.e., the geodesic,
$x^\mu(\tau)$ ($\mu=0,...,3$) are particle coordinates,
and $g_{\mu\nu}(x)$ is a metric tensor in a chosen chart of the space-time
considered.
The action \re{2.1} is invariant with respect to an arbitrary change of the evolution parameter: $\tau\to\tau'=f(\tau)$,
since the parametrization of geodesics has no physical meaning.
For the de Sitter space-time \cite{dS17} geodesics were studied
from different viewpoints \cite{dS17,2017MPLA...3250223C,2018MPLA...3375002C,CGKM08}
in many coordinate charts introduced for this space-time
\cite{CGKM08,0904.4184,1608.02792}.

It is convenient to consider de Sitter space-time as a 4-dimensional
hyperboloid $\Bbb H$:
%
\begin{equation}\lab{2.2}
\eta_{MN} y^M y^N:=(y^0)^2-(y^1)^2-\dots-(y^4)^2=-R^2
\end{equation}
in 5-dimensional Minkowski space ${\Bbb M}_5$ with coordinates
$y^M$ ($M=0,1,\dots,4$) and the metrics  $||\eta_{MN}||={\rm
diag}(+,-,\dots,-)$;\  \cite{Cox43,CGKM08}. The
constant $R$ determines the scalar curvature ${\cal R}$ of the
de~Sitter space, and it is related with the cosmological
$\Lambda$-constant: ${\cal R}=12/R^2=4\Lambda$; the speed of light
is put $c=1$.

The hyperboloid $\Bbb H$ is invariant with respect to de Sitter
group O(1,4) represented in ${\Bbb M}_5$ by standard linear
pseudoorthogonal transformations. Thus we will use
standard notations for O(1,4)-invariants $y\cdot z:=\eta_{MN}y^Mz^N$ and $y^2:=y\cdot y$
built of arbitrary 5-vectors $y$, $z\in{\Bbb M}_5$.

The embedding ${\Bbb H}\hookrightarrow{\Bbb M}_5$ implies, in terms of local coordinates $x^\mu$ in de Sitter space,
a set of appropriate functions $y^M(x)$ turning the equation \re{2.2} into identity \cite{CGKM08,0904.4184,1608.02792}.
Then the pseudo-Euclidian O(1,4)-invariant metrics is pulled back naturally from ${\Bbb M}_5$ onto ${\Bbb H}$:
%
\begin{equation}\lab{2.7}
\rho(x, x'):= \left.(y-y')^2\right|_{\Bbb H}.
\end{equation}
This endows the de Sitter space with a causal structure of the ambient Minkowski space:
\begin{itemize}
\item
the interval between points $x, x'\in{\Bbb H}$ is timelike if $\rho(x, x')>0$, i.e.,
if $y'\in{\Bbb H}\subset{\Bbb M}_5$ lies inside the light cone with a vertex $y\in{\Bbb H}\subset{\Bbb M}_5$ (or the same with
$y$ and $y'$ permuted);
\item
the interval is spacelike if $\rho(x, x')<0$, i.e.,
if $y'$ lies outside the light cone;
\item
the interval is isotropic if $\rho(x, x')=0$, i.e.,
if $y'$ lies on the light cone hypersurface.
\end{itemize}
For infinitely closed 5-vectors $y$ and $y'=y+\D{}{y}$ the function \re{2.7} yields
the pseudo-Riemannian metrics involved in the action integral \re{2.1} for the case of de Sitter space:
%
\begin{equation*}
\D{}{s^2}:=\left.\eta_{MN}\D{}{ y^M}\D{}{ y^N}\right|_{\Bbb H}=
g_{\mu\nu}(x)\D{}{ x}^\mu\D{}{ x}^\nu.
\end{equation*}

Thus the test particle dynamics in de
Sitter space can be reformulated to some variational principle
with a constraint, defined in the configuration space ${\Bbb M}_5$
\cite{AAMP07,Cas08,CGKM08}. The simplest version is \cite{CGKM08}:
%
\begin{equation}\lab{2.4}
I=-\int\D{}{\tau}\left\{m\sqrt{\dot y^2(\tau)}-\lambda(\tau)( y^2(\tau)+R^2)\right\},
\end{equation}
where the condition \re{2.2} is taken into account as a holonomic constraint
by means of the Lagrange multiplier $\lambda(\tau)$. The Euler-Lagrange
equation for 5-vector $y(\tau)$ representing the particle position $x(\tau)\in{\Bbb H}$
can be written down in the following manifestly covariant form
%
\begin{equation}\lab{2.5}
\frac{\D{}{ }}{\D{}{\tau}}\frac{\dot y}{\sqrt{\dot y^2}}-\sqrt{\dot
y^2}\frac{y}{R^2}=0
\end{equation}
which is invariant with respect to both the O(1,4) group and an
arbitrary change of the evolution parameter $\tau$.
The solution of the geodesic equation \re{2.5} is
%
\begin{equation}
y(\tau)=y(0)\cosh\frac{s(\tau)}R+R\frac{\dot y(0)}{\sqrt{\dot y^2(0)}}\sinh\frac{s(\tau)}R,
\lab{2.6}\\
\end{equation}
where the constant 5-vectors $y(0)$ and $\dot y(0)$ are subjected
to the constraint \re{2.2} and its differential consequence $y\cdot\dot
y=0$, and $s(\tau)$ is the proper time elapsed from $y(0)$ to $y(\tau)$:
%
\begin{equation}
s(\tau):=\int\nolimits_{0}^{\tau}\!\!\D{}{\tau}\sqrt{\dot y^2(\tau)}.
\lab{2.6a}
\end{equation}
The proper time as a function of $\tau$ cannot be determined from the equation \re{2.5}, due to reparametrization invariance,
but it can be chosen by hands for a conveniency. For example, with the proper time parametrization $s(\tau):=\tau$
we have $\dot y^2=1$, and the equation \re{2.6} reproduces the de Sitter geodesic found in Ref. \cite{CGKM08}.

Due to de Sitter symmetry, there exists 10 integrals of motion
collected in the skew-symmetric angular 5-momentum tensor:
%
\begin{equation}\lab{2.8}
J_{MN}= y_M\pi_N-y_N\pi_M= -J_{NM},
\end{equation}
where
%
\begin{equation}\lab{2.9}
\pi_M=m\dot y_M/\sqrt{\dot y^2}
\end{equation}
are components of 5-momentum.

At this point one can develop the covariant Hamiltonian description
on the phase space ${\rm T}^*{\Bbb M}_5$ with variables $y^M$,
$\pi_N$ ($M,N=0,...,4$) and standard Poisson brackets:
$\{y^M,y^N\}=0$, $\{\pi_M,\pi_N\}=0$, $\{y^M,\pi_N\}=\delta^M_N$.
The integrals of motion $J_{MN}$ become canonical generators of
O(1,4) group while the Legendre transform \re{2.9} is degenerated
due to the reparametrization invariance of the action \re{2.4}. Thus the
canonical Hamiltonian vanishes while the {\em mass-shell} constraint
arises, $\pi^2-m^2=0$, apart to the holonomic constraint \re{2.2}.
Both constrains are primary ones according to Dirac's terminology of canonical formalism with constraints
\cite{Dir64}. They form the primary Dirac's Hamiltonian: $H'_{\rm D} = \lambda(\pi^2-m^2)+\lambda_1(y^2+R^2)$,
where $\lambda$ and $\lambda_1$ are Lagrange multipliers.
The compatibility conditions
%
\begin{eqnarray*}
\{y^2+R^2,H'_{\rm D}\}&=&4\lambda y\cdot\pi\approx0, \qquad \{\pi^2-m^2,H'_{\rm D}\}=-4\lambda_1y\cdot\pi\approx0,
\end{eqnarray*}
give rise to the secondary constraint
$
y\cdot\pi=0
$
so that Dirac's Hamiltonian at this stage takes the form:
$H''_{\rm D} = H'_{\rm D}+\lambda_2\,y\cdot\pi$.
Reexamining compatibility conditions gives no new constraints but fixes
partially Lagrange multipliers: $\lambda_1=0$. Putting then $\lambda_2=-y\cdot\pi/y^2$
yields the final Dirac's Hamiltonian $H_{\rm D} = \lambda(\tau)\phi(y,\pi)$ with the unspecified Lagrange multiplier $\lambda(\tau)$
(due to the reparametrization invariance) and the function $\phi(y,\pi)$ which determines the modified mass-shell constraint
%
\begin{equation}\lab{2.12}
\phi:= \pi_{\bot}^2-m^2\equiv \ha J^2/y^2-m^2=0;
\end{equation}
here $\displaystyle{\pi_{\bot M}:= \pi_M
-\frac{y\cdot\pi}{y^2}y_M \approx \frac{J_M^{\ N}y_N}{R^2}}$ (so
that $y\cdot\pi_\bot\equiv0$) and $J^2:= J_{MN}J^{MN}$.
The symbol ``~$\approx$~'' denotes a ``weak equality'', i.e. by virtue of the holonomic constraint  \re{2.2}; \cite{Dir64}.

Let us note that the set of constraints \re{2.2} and \re{2.12} are
the 1st class \cite{Dir64}, i.e., they satisfy the identity:
$\{y^2+R^2,\phi\}\equiv0$. Together with the Dirac's Hamiltonian
$H_{\rm D} = \lambda\phi$ these constraints endow effectively the
system with three degrees of freedom (as it should). Henceforth the
quantity $y\cdot\pi$ is not involved in the dynamics, and the
secondary constraint $y\cdot\pi=0$ can be abandoned.

The Hamiltonian equation for the particle position 5-vector $y$ reads:
%
\begin{equation}\lab{2.15}
\dot
y=\lambda\{y,\phi\}=2\lambda\pi_\bot\approx\frac{2\lambda}{R^2}{\s J}\,y.
\end{equation}
Note that the matrix $\s J:=||J^M_{\ \ N}||:=||\eta^{ML}J_{LN}||$ is
conserved, thus the equation \re{2.15} is linear. Its formal
solution follows immediately: $y(\tau)={\rm
e}^{\frac{s(\tau)}{mR^2}{\s J}}y(0)$, where the unspecified function
$s(\tau)=2m\int_0^\tau\D{}{\tau}\lambda(\tau)$ is the Hamiltonian
image for the proper time function \re{2.6a}. The Cauchy problem
becomes solved after the matrix $\s J$ is expressed in terms of
initial values $y(0)$ and $\dot y(0)$ by the equalities \re{2.8},
\re{2.9} and their consequences ${\s J}y\approx mR^2v$, ${\s
J}v=my$, where $v=\dot y/\sqrt{\dot y^2}$. Then expanding the
exponent in power series reproduces the solution \re{2.6}.

It may seem unreasonable the use of 5-dimensional reparametrization invariant
description together with Dirac's formalism with constraints in order to
derive geodesics in de Sitter space. These tools, however, appear effective when considering
two-body problems in next sections.


\section{Action-at-a-distance dynamics of two particles in de Sitter space}
\renewcommand{\theequation}{3.\arabic{equation}}
\setcounter{equation}{0}

In the framework of Wheeler-Feynman electrodynamics
\cite{Roh90,YaT12,H-N74,V-T86} a system of charged point-like
particles is described by the Tetrode-Fokker action-at-a-distance
variational principle \cite{Tet22,Fok29}. This formalism was
generalized for a curved space-time by Hoyle and Narlikar
\cite{H-N74} and others \cite{V-T86,Tur86}.

For the system of two charged particles of masses $m_a$ and charges
$e_a$ ($a=1,2$) the Tetrode-Fokker action integral has a form:
%
\begin{eqnarray}
\lab{3.1} I&=&I_{\rm free}+I_{\rm int},\quad \mbox{where}\quad
I_{\rm free}=-\sum_{a=1}^2
m_a\inta\D{}{s_a},\qquad\\
\lab{3.1a} &&\D{}{s_a}:=\sqrt{g_{\mu\nu}(x_a(\tau_a))\dot{x}_a^\mu(\tau_a)\dot{x}_a^\nu(\tau_a)}\,\D{}{\tau_a},\\
\lab{3.2} &&I_{\rm int}= -4\pi
e_1e_2\intab\D{}{x_1^\mu}\D{}{x_2^\nu}G_{\mu\nu}(x_1,x_2);
\end{eqnarray}
here $x_a^\mu(\tau_a)$ ($\mu=0,...,3$) are space-time coordinates of
particle  world lines parameterized by evolution parameters $\tau_a$
($a=1,2$). Free-motion terms $I_{\rm free}$ of the action \re{3.1}
have the form \re{2.1} for each particle. An integrand of the
interaction term \re{3.2} is the symmetric Green function
$G_{\mu\nu'}(x,x')$ of
the covariant wave equation $\square A_\mu+{{\cal R}_\mu}^\nu A_\nu=0$
for the electromagnetic potential $A_\mu$ \cite{PPV11,Nar70};
here $\square$ is the d'Alembertian in a curved space-time considered,
and ${{\cal R}_\mu}^\nu$ is the Ricci tensor.
For the curved space-time $G_{\mu\nu'}(x,x')$ is a bi-vector function
which construction in general is a complicated problem \cite{PPV11}.

For de Sitter space-time the symmetric Green function is known from
Ref. \cite{H-Ch08}\footnote{An earlier proposal \cite{Nar70}
is unappropriate as it does not meet demands of de
Sitter-covariance.}. It is presented here in geometric terms which are
indifferent to a choice of coordinate chart:
%
\begin{eqnarray}
G_{\mu\nu'}(x,x')&=&G^{\delta}_{\mu\nu'}(x,x')+G^{\Theta}_{\mu\nu'}(x,x');
\lab{3.3}
\end{eqnarray}
here
\begin{eqnarray}
G^{\delta}_{\mu\nu'}(x,x')&:=&\frac1{16\pi }\bar
g_{\mu\nu'}(x,x')\,\delta(\rho(x,x'))
\lab{3.4}\\
G^{\Theta}_{\mu\nu'}(x,x')&:=&-\frac1{24\pi
R^2}\left\{\left(\frac1Z+\frac1{2Z^2}\right)\bar g_{\mu\nu'}
+\frac{R^2}{Z^3}(\partial_\mu
Z)(\partial_{\nu'}Z)\right\}\Theta(\rho(x,x'));
\lab{3.5}\\
\bar
g_{\mu\nu'}(x,x')&:=&-2R^2\left\{\partial_\mu\partial_{\nu'}Z-\frac{1}{Z}(\partial_\mu
Z)(\partial_{\nu'}Z)\right\},
\lab{3.6}\\
\quad Z(x,x')&:=&1+\qu\rho(x,x')/R^2, \lab{3.7}
\end{eqnarray}
where $\bar g_{\mu\nu'}(x,x')$ is the parallel propagator \cite{PPV11,Nar70},
and the metric function $\rho(x,x')$ is defined by \re{2.7}.
We note that the Green function \re{3.3}
consists of two parts. The local part
\re{3.4} is proportional to the Dirac $\delta$-function and thus
supported by the light cone surface $\rho(x,x')=0$. The non-local part
\re{3.5} is  proportional to the Heaviside $\Theta$-function and
thus supported by the light cone interior $\rho(x,x')>0$. This is a
common feature of curved space-times \cite{PPV11}, contrary to the Minkowski
space-time, where Green functions of massless fields have a local
part only. But in present case of de Sitter space-time the non-local
contribution \re{3.5} of the Green function \re{3.3} in the integral \re{3.2} can
be effectively reduced to a local one \cite{D-Y19E}.

In order to show this let us first introduce the relative position 5-vector
$r\equiv y_1-y_2$, the particle unit 5-velocities $v_a\equiv\dot y_a/\sqrt{\dot y_a^2}$,
and the dimensionless scalars of these 5-vectors $v_1\cdot v_2$ and $r\cdot v_a/R$ ($a=1,2$)
which are homogeneous functions of degree zero of derivatives $\dot y_1$ and $\dot y_2$.
It is convenient for a subsequent interim calculation to present these scalars as follows:
%
\begin{equation}\lab{3.8}
\omega:= v_1\cdot v_2|_{\Bbb H}=-\frac{1}{2}\frac{\D{2}{\rho(x_1,x_2)}}{\D{}{s_1}\D{}{s_2}},\qquad
\nu_a:= \left.\frac{r\cdot v_a}R\right|_{\Bbb H}=-\frac{(-)^a}{2R}\frac{\D{}{\rho(x_1,x_2)}}{\D{}{s_a}},\quad
a=1,2,
\end{equation}
where the function $\rho(x_1,x_2)$ and the interval elements $\D{}{s_a}$ are defined by
eqs. \re{2.7} and \re{3.1a}, respectively. Note that the differentiation over $\D{}{s_1}$
(or $\D{}{s_2}$) acts on $x_1(\tau_1)$ (or $x_2(\tau_2)$).

In these terms the integrand of the interaction term \re{3.2} of the action \re{3.1} reads:
%
\begin{eqnarray*}
\D{}{x_1^\mu}\D{}{x_2^\nu}G_{\mu\nu}(x_1,x_2)=\frac{\D{}{s_1}\D{}{s_2}}{4\pi}
\left\{\left(\omega-\frac{\nu_1\nu_2}{2Z}\right)\delta(\rho)
- \left(\frac{2Z+1}{Z^2}\omega-\frac{Z+1}{Z^3}\nu_1\nu_2\right)\frac{\Theta(\rho)}{12
R^2}\right\}.&&
\end{eqnarray*}
Then, applying the integration-by-part formula:
%
\begin{eqnarray*}
\int\limits_{-\infty}^{+\infty}
\int\limits_{-\infty}^{+\infty} \D{}{s_1}\D{}{s_2}\, \omega
F(\rho)=-\frac12\int \limits_{-\infty}^{+\infty}
\int\limits_{-\infty}^{+\infty} \D{}{s_1}\D{}{s_2}\,
\frac{\D{2}{\rho}}{\D{}{s_1}\D{}{s_2}}F(\rho) \nonumber \\
=-2R^2\int\limits_{-\infty}^{+\infty} \int\limits_{-\infty}^{+\infty}
\D{}{s_1}\D{}{s_2}\, \nu_1\nu_2 \, \frac{\D{}{F(\rho)}}{\D{}{\rho}}
-\left.\frac12\,\rho F(\rho)
\right|_{s_1=-\infty\atop s_2=-\infty}^{s_1=+\infty\atop s_2=+\infty},
\end{eqnarray*}
which holds for any function $F(\rho)$, to the Tetrode-Fokker
integral \re{3.2}, one obtains:
%
\begin{equation}\lab{3.10}
I_{\rm int}= -4\pi
e_1e_2\intab\D{}{\tau_1}\D{}{\tau_2}\,\dot x_1^\mu\dot x_1^\nu \, G_{\mu\nu}(x_1,x_2)
\simeq -e_1e_2\intab\D{}{s_1}\D{}{s_2}\,\omega\delta(\rho),
\end{equation}
where the symbol ``~$\simeq$~'' denotes an equality up to boundary
terms which do not contribute in variational problem.

It is remarkable that the only local (i.e., light cone surface)
contribution of Green function remains in the Tetrode-Fokker
integral \re{3.10}; this structure is a necessary starting point for
a construction of the model of Staruszhkiewicz-Rudd-Hill type in
next Section.

Similarly, one can consider a particle system with the scalar
interaction. The interaction term of the Fokker-type action \re{3.1}
in this case has a form \cite{H-N74}:
%
\begin{equation}\lab{3.11}
I_{\rm int}= -4\pi g_1g_2\intab\D{}{s_1}\D{}{s_2}\,G(x_1,x_2),
\end{equation}
where $g_a$ ($a=1,2$) are scalar ``charges'' of particles, and
the bi-scalar function $G(x,x')$ is the symmetric Green function of the wave equation
$\square \varphi=0$ for a scalar field $\varphi$
mediating the interaction and {\em minimally} coupled to a
gravitation \cite{PPV11}. For the de Sitter space-time the Green function
$G(x,x')$ was found by Narlikar \cite{Nar70}:
%
\begin{eqnarray}
G(x,x')&=&G^{\delta}(x,x')+G^{\Theta}(x,x'):=\frac1{4\pi}\left\{\delta(\rho)-\frac1{2R^2}\Theta(\rho)\right\}.
\lab{3.12}
\end{eqnarray}
In contrast to the case of electromagnetic interaction, the nonlocal
contribution $G^{\Theta}(x,x')$ of the Green function \re{3.12} is
essential: it cannot be removed from the action \re{3.11} by means
of the integration by parts or another equivalent transformation.

The Penrose-Chernikov-Tagirov equation $(\square - {\cal R}/6)\varphi=0$
corresponds to a {\em conformal} coupling of the scalar field to a
gravitation \cite{Pen64,Ch-T68}.
In the case of de Sitter space-time the
scalar curvature ${\cal R}=12/R^2$ is constant, and the Green
function can be found easily by means of distributional methods
\cite{PPV11}. It appears purely local:
%
\begin{equation}\lab{3.13}
G(x,x')=G^{\delta}(x,x'):=\frac1{4\pi}\delta(\rho).
\end{equation}

The electromagnetic \re{3.10} and scalar \re{3.11}, \re{3.13}
interaction terms of the Fokker-type action admit the obvious
de-Sitter-invariant generalization:
%
\begin{equation}\lab{3.14}
I_{\rm int}=
-\intab\D{}{s_1}\D{}{s_2}\,f(\nu_1,\nu_2,\omega)\delta(\rho),
\end{equation}
where $\D{}{s_a}$ are defined in \re{3.1a}, and $f(\nu_1,\nu_2,\omega)$
may be an arbitrary function of its three scalar arguments \re{3.8}, so it is a
homogeneous function of degree zero of $\dot y_1$ and $\dot y_2$.
Thus the expression \re{3.14} possesses both the de Sitter  invariance
and the double reparametrization invariance.
It comprises a variety of interactions which may have a field-theoretical nature or can be
introduced phenomenologically.


\section{Time-asymmetric models in de Sitter space-time}
\renewcommand{\theequation}{4.\arabic{equation}}
\setcounter{equation}{0}

Staruszkiewicz \cite{Sta70,Sta71}, Rudd and Hill \cite{R-H70}
replaced in the Tetrode-Fokker action integral the symmetric Green
function $G$ of d'Alembert equation by the retarded $G^{(+)}$ or
advanced  $G^{(-)}$ Green function:
$G^{(\pm)}(x_1,x_2)=2\Theta[\pm(x_1^0-x_2^0)]G(x_1,x_2)$. This have
led them to a two-particle model with the time-asymmetric
retarded-advanced interaction. Following this idea, one should
insert in the general interaction term \re{3.14} of the Fokker-type
action \re{3.1} the factor
$2\Theta[\eta(x_1^0-x_2^0)]=2\Theta[\eta(y_1^0-y_2^0)]$, where
$\eta$ = +1 or --1. Then, similarly to the singe-particle case considered
in Section 2,
it is convenient to present this Fokker-type action
via global variables in the ambient Minkowski space ${\Bbb M}_5$.
One thus obtains:
%
%
\begin{eqnarray}\lab{4.2}
I_{\rm int}&=& -\intab\D{}{\tau_1}\D{}{\tau_2}\sqrt{\dot
y_{\smash1}^2}\sqrt{\dot
y_{\smash2}^2}\,f(\nu_1,\nu_2,\omega)\,2\Theta(\eta
r^0)\,\delta(r^2)|_{{\Bbb H}^2},
\end{eqnarray}
where the integrand in r.-h.s. of \re{4.2} is constrained on
${\Bbb H}^2={\Bbb H}\times{\Bbb H}$, i.e., the particle position 5-vectors
$y_a(\tau_a)$ ($a=1,2$) are subjected to the {\em hyperboloid} conditions for each particle:
%
\begin{equation}\lab{4.1}
y_a^2+R^2=0,\qquad a=1,2.
\end{equation}

An integrand of the double integral $I_{\rm int}$ in \re{4.2} is non-zero provided
%
\begin{equation}\lab{4.3}
r^2:=(y_1-y_2)^2 = 0 , \qquad \eta r^0:= \eta(y_1^0-y_2^0)>0.
\end{equation}
This condition can be treated as the equation of past or future {\em light cone}, depending on the value
$\eta=\pm1$ and the choice which point $y_1$ or $y_2 $ is a vertex of the cone.
If the time-symmetric action \re{3.14} is invariant under a particle permutation,
the invariance of the corresponding time-asymmetric action \re{4.2} is provided by the additional change $\eta\to-\eta$.

From a physical viewpoint, the choice of the sign factor $\eta=\pm1$ is indifferent. Both cases correspond to the electromagnetic
interaction with a ``spoiled'' causality. They lead to distinguished two-body problems which differ from one another and from those
of Wheeler-Feynman or retarded electrodynamics. It is worth noting that in the case of the flat-space Staruszkiewucz-Rudd-Hill model
the particle world lines corresponding to different $\eta=\pm1$ are distinguishable only in a highly  relativistic domain
\cite{Sta70,Duv98}.

The Fokker-type action integral \re{3.1}, \re{4.2} is invariant with respect to an arbitrary change of
each parameter $\tau_1$ and $\tau_2$. Thus two of ten variables
$y_1^M(\tau_1)$, $y_2^M(\tau_2)$ ($M=0,...,4$) to be found remain undetermined within the variational problem.
It is profitable to fix partially this functional arbitrariness by hands as follows.
Let us choose one of variables, say $y_2^0(\tau_2)$, in such a way that the condition \re{4.3} turns into identity at
$\tau_1=\tau_2$. This implies that both particle world lines are parameterized by a common
evolution parameter, say $\tau_1$, and the simultaneous events $y_1(\tau_1)$ and  $y_2(\tau_1)$ lie on the
isotropic light cone surface \re{4.3}. Using the equality (see \cite{Sta71})
%
\begin{eqnarray*}
2\,\Theta\Bigl[\eta\Bigl(y_1^0(\tau_1) -
y_2^0(\tau_2)\Bigr)\Bigr]
\,\delta\!\left[\Bigl(y_1(\tau_1) - y_2(\tau_2)\Bigr)^2\right]
=\frac{\delta(\tau_1-\tau_2)}{
  \Bigl|\dot y_2(\tau_2)\cdot\Bigl(y_1(\tau_1) -
         y_2(\tau_2)\Bigr)\Bigr|}
\end{eqnarray*}
in the interaction term  \re{4.2}
and integrating over $\tau_2$ reduces
the Fokker-type action \re{3.1} to the single-time form
%
\begin{equation}\lab{4.5a}
I=\int\D{}{\tau}\,\tilde L
\end{equation}
with the lagrangian $\tilde L:=L|_{{\mathrm T}{\Bbb K}}$, where
%
\begin{equation} \label{4.5}
L \ =\ -\sum_{a=1}^2 m_a \sqrt{\dot y_a^2} \ - \ \sqrt{\dot
y_{\smash1}^2}\sqrt{\dot
y_{\smash2}^2}\frac{f(\nu_1,\nu_2,\omega)}{|\dot y_2 \cdot r|}.
\end{equation}
The Lagrangian $\tilde L$ is defined on the tangle bundle ${\mathrm T}{\Bbb K}$
over the 7-dimensional configuration manifold
${\Bbb K}\subset{\Bbb H}^2\subset{\Bbb M}_5^2\equiv{\Bbb M}_5\times{\Bbb M}_5$
described by the conditions \re{4.1}, \re{4.3}. The corresponding variational
problem gives rise to second order differential equations of motion and thus the
transition to the usual Hamiltonian description is straightforward.

The Lagrangian $\tilde L$ (as well as $L$) is the first degree
homogeneous function of particle velocities. Thus the action
\re{4.5a} possesses a residual invariance with respect to an
arbitrary change of the common evolution parameter: $\tau$.
This symmetry allows one to fix a remaining timelike variable
by hands and, together with the conditions \re{4.1}, \re{4.3}, enables to
arrive at the ordinary Lagrangian description in the 6-dimensional configuration space ${\Bbb Q}$.
In practice, however, an explicit elimination of redundant variables,
say $y_1^0$, $y_1^4$, $y_2^0$, $y_2^4$, breaks a manifest 5-dimensional Lorentz-covariance,
and makes a subsequent treatment cumbersome. As usual, a success in solving equations of motion
is predetermined by an appropriate parametrization
of the configuration space which is not evident in the case  of ${\Bbb Q}$.

    An alternative way is the use of a manifestly covariant Lagrangian
description in the 10-dimensional configuration space ${\Bbb M}_5^2$.
In this case an unconditional extremum problem of the action \re{4.5a} is modified in favor
of an equivalent conditional extremum problem of
%
\begin{equation}\lab{4.4}
I = \inta\D{}{\tau}\,\left\{L + \lambda_0 r^2+\sum_{a=1}^2\lambda_a (y_a^2+R^2)\right\}
\end{equation}
with the Lagrangian function \re{4.5} defined on ${\mathrm T}{\Bbb M}_5^2$. The Lagrangian
multipliers $\lambda_0(\tau)$, $\lambda_a(\tau)$ take the conditions \re{4.3}, \re{4.1} into account
as holonomic constraints; the unilateral constraint $\eta r^0>0$ is implied as well.

de Sitter invariance of the Lagrangian \re{4.5} and constraints
\re{4.1}, \re{4.3} provides the existence of ten Noether integrals
of motion, collected in the angular 5-momentum tensor:
%
\begin{eqnarray}
\label{4.14} J_{MN} = \sum^2_{a=1} \left(y_{aM}\pi_{aN} -
y_{aN}\pi_{aM}\right),
\end{eqnarray}
where \vspace{-5ex}\\
%
\begin{eqnarray}
\label{4.15} \pi_{aM} = \partial L/\partial\dot y^{M}_a,\qquad
a=1,2.
\end{eqnarray}
Besides,  the Lagrangian \re{4.5} satisfies the identity:
%
\begin{equation} \label{4.16}
\sum_{a=1}^2\dot y_a\!\cdot \!\pi_a  - L=0,
\end{equation}
due to the reparametrization invariance of the action \re{4.4}.


\section{Canonical formalism with constraints}
\renewcommand{\theequation}{5.\arabic{equation}}
\setcounter{equation}{0}

    The Lagrangian description in the configuration space ${\Bbb M}_5^2$
enables a natural transition to the manifestly covariant Hamiltonian
description with constraints \cite{Dir64}. The corresponding
20-dimensional phase space ${\rm T}^*{\Bbb M}_5^2$
with the particle canonical variables $y_a^M$, $\pi_{bN}$
($a,b=1,2$; $M,N=0,\dots,4$) is endowed with
the standard Poisson brackets: $\{y_a^M,y_b^N\}=0$,
$\{\pi_{aM},\pi_{bN}\}=0$,
$\{y_a^M,\pi_{bN}\}=\delta_{ab}\delta^M_N$.

    Components of the conserved angular 5-momentum tensor \re{4.14} become,
within the Hamiltonian description, the generators $J_{MN}$ of the
canonical realization of the de Sitter group, i.e., they satisfy the
canonical relations of the Lie algebra of O(1,4):
%
\begin{eqnarray}\label{5.1}
&&\{J_{MN}, J_{LK}\}=\eta_{ML}J_{NK} + \eta_{NL}J_{ML} -
\eta_{MK}J_{NL} - \eta_{NL}J_{MK}.
\end{eqnarray}

    Due to the identity \re{4.16} the Legendre transformation \re{4.15} is degenerated,
the canonical Hamiltonian vanishes, while the additional constraint
arises \cite{Dir64}, similarly to the mass-shell constraint in the single particle case.
The function determining this constraint constitutes (together with the holonomic constraints \re{4.1}, \re{4.3})
a primary Dirac's Hamiltonian.

The subsequent procedure is similar to that of the single particle case in Section 2.
The compatibility conditions of the dynamics with primary constraints give rise to secondary constraints
which then are combined with the primary constraints in the secondary Dirac' Hamiltonian etc.
In a final compatible form the dynamics is generated by the Dirac's Hamiltonian
$H_{\rm D} = \lambda(\tau)\Phi(y_a,\pi_b)$ where $\lambda(\tau)$ is unspecified
Lagrange multiplier (due to the reparametrization invariance), and the constraint
$\Phi(y_a,\pi_b)=0$ is the first class with respect to the holonomic constraints \re{4.1}, \re{4.3},
i.e., the function $\Phi(y_a,\pi_b)$ satisfies the equalities:
%
\begin{eqnarray}\label{5.2}
\{\Phi,r^2\}=0, \qquad \{\Phi,y^2_a +R^2\}=0,\quad a=1,2.
\end{eqnarray}
Besides, this constraint must be de Sitter invariant asinmuch the angular momentum tensor
\re{4.14} must be conserved.

We will refer to $\Phi(y_a,\pi_b)=0$ as the {\em dynamical} constraint for two reasons.
Firstly, the function $\Phi(y_a,\pi_b)$ generates an evolution via Dirac's Hamiltonian.
Secondly, a specific form of $\Phi(y_a,\pi_b)$ is determined by the Lagrangian \re{4.5}, in particular, by the form of
the interaction function $f(\nu_1,\nu_2,\omega)$ chosen. However, the equations \re{5.2} and
de Sitter invariance requirements are sufficient to outline a general structure of the
dynamical constraint and the corresponding Hamiltonian mechanics.

Let functions of canonical variables $\varphi(y_a,\pi_b)$ which satisfy the conditions \re{5.2}
be referred to as {\em observables} in Dirac's meaning \cite{Dir64}.
We will use sometimes the collective canonical variables:
%
\begin{eqnarray}\label{5.0}
\hspace{-2ex}
Y^M=\ha(y^M_1+y^M_2),\quad r^M=y^M_1-y^M_2,\quad
\Pi_{M}=\pi_{1M}+\pi_{2M},\quad \pi_{M}=\ha(\pi_{1M}-\pi_{2M}).\quad
\end{eqnarray}
The components of position 5-vectors $Y$, $r$ are the observables.
Solving the equations \re{5.2} yields other observables, the momentum-type 5-vectors $\Pi_{\bot}$,
$\pi_{\bot}$ with the components:
%
\begin{eqnarray}
\Pi_{\bot M}&:=&\frac{Y^L
J_{LM}}{Y^2}\approx\Pi_M+\frac{(Y\cdot\Pi)Y_M+(Y\cdot\pi)r_M}{R^2},
\label{5.3}\\
\pi_{\bot
M}&:=&\left(\delta_M^N-\frac{r_M\Pi_\bot^N}{\Pi_\bot\!\cdot
r}\right) \left(\delta_N^L-\frac{Y_NY^L}{Y^2}\right)\pi_L
\label{5.4}
\end{eqnarray}
which are not all independent due to the identities:
$\Pi_{\bot}\!\cdot Y\equiv0$, $\pi_{\bot}\!\cdot
Y\equiv0$, $\Pi_{\bot}\!\cdot \pi_{\bot}\equiv0$.

A set of functions $\varphi(Y, r, \Pi_{\bot}, \pi_{\bot})$
constitutes a complete algebra of observables which is
closed with respect to Poisson brackets.
Indeed, if $\varphi_1$ and $\varphi_2$ are observables then
$\{\varphi_1,\varphi_2\}$ is observable due to the Jacobi identity.
The particle positions $y_a$ and the dynamical constraint $\Phi(Y, r, \Pi_{\bot}, \pi_{\bot})$
are observable, thus the algebra of observables is sufficient to formulate
equations of motion generated by the Dirac's Hamiltonian $H_{\rm D} = \lambda\Phi$.
Aforementioned secondary constraints are not observable and can be abandoned,
similarly to the single particle case.

The requirement of the dynamical constraint to be de Sitter invariant
yields the following its general structure:
%
\begin{equation} \label{5.5}
\Phi(\Pi_\bot^2,~\pi_\bot^2, ~\Pi\!\cdot\!r, ~\pi\!\cdot\!r) = 0,
\end{equation}
where $\Phi$ may be an arbitrary function of its scalar arguments:
$\Pi_\bot^2,~\pi_\bot^2$ and
$\Pi\cdot r\approx\Pi_\bot\cdot r$,
~$\pi\cdot r\approx\pi_\bot\cdot r$;
here the use of weak equality ``$\,\approx\,$'' by virtue of the
holonomic constraints \re{3.1}, \re{3.3} simplifies the dynamical constraint but
does not affects the dynamics of observables.

The dynamical constraint \re{5.5} determines implicitly one of the argument
of $\Phi$ as a function of three other arguments.
Since this function can be regarded arbitrary within a general consideration,
the Hamiltonian formalism with constraints \re{4.1}, \re{4.3}, \re{5.5}
embraces a variety of two-particle systems as wide as the
original Fokker-type or Lagrangian formalism with the arbitrary function  $f(\nu_1,\nu_2,\omega)$ does,
except, perhaps, some special cases.

The procedure how to obtain the dynamical constraint, given the
interaction function $f(\nu_1,\nu_2,\omega)$ in the Lagrangian \re{4.2},
is described in detail in Appendix A. In practice, however,
elementary algebraic operations
implied there can be rarely finished in a closed form.

Fortunately, two physically motivated examples, considered in
Section 3, are the cases. For the electromagnetic time-asymmetric
interaction one puts  in \re{4.5} $f=\alpha_{\rm e}\,\omega$, where
$\alpha_{\rm e}:= e_1e_2$, and arrives at the following dynamical
constraint:
%
\begin{eqnarray} \label{5.12}
\Phi_{\rm e}:=\pi_{\bot}^2 +\qu\Pi_\bot^2
+\frac1{4R^2}\left(\frac{(\pi_1\!\cdot\!r)(\pi_2\!\cdot\!r)}{\eta\Pi\!\cdot\!r}
- \alpha_{\rm e}\right) \left(\eta\Pi\!\cdot\!r-4\alpha_{\rm
e}\right)-\frac{m_1^2\,\pi_2\!\cdot\!r+m_2^2\,\pi_1\!\cdot\!r}{\Pi\!\cdot\!
r}
\nn\\
{}-\alpha_{\rm e}\frac{\Pi_\bot^2-m_1^2-m_2^2}{\eta\Pi\!\cdot\!r}
+\alpha_{\rm e}^2\left(\frac{m_1^2}{\eta\pi_1\!\cdot\!r-\alpha_{\rm
e}}+\frac{m_2^2}{\eta\pi_2\!\cdot\!r-\alpha_{\rm e}}\right) =0.
\end{eqnarray}
For the scalar interaction $f=\alpha_{\rm s}:= g_1g_2$, and the dynamical constraint has the form:
%
\begin{eqnarray} \label{5.13}
\Phi_{\rm s}:=\pi_{\bot}^2 +\qu\Pi_\bot^2
+\frac{(\pi_1\!\cdot\!r)(\pi_2\!\cdot\!r)}{4R^2}-\frac{m_1^2\,\pi_2\!\cdot\!r+m_2^2\,\pi_1\!\cdot\!r
+2\eta\alpha_{\rm s}m_1m_2}{\Pi\!\cdot\!
r\left(1-\displaystyle{\frac{\alpha_{\rm
s}^2}{(\pi_1\!\cdot\!r)(\pi_2\!\cdot\!r)}}\right)} =0.
\end{eqnarray}
Both the constraints \re{5.12} and \re{5.13} reduce in the free-particle limit
$\alpha\to0$ to the constraint:
%
\begin{eqnarray}
\Phi_{\rm free}&:=&\pi_{\bot}^2 +\qu\Pi_\bot^2
+\frac{(\pi_1\!\cdot\!r)(\pi_2\!\cdot\!r)}{4R^2}
-\frac{m_1^2\,\pi_2\!\cdot\!r+m_2^2\,\pi_1\!\cdot\!y}{\Pi\!\cdot\!
r}.
\label{C.1a}
\end{eqnarray}
This case of the time-asymmetric system with no interaction (i.e., $f=0$)
deserves a particular consideration.
The free-particle dynamical constraint \re{C.1a} is not the additive function in variables of
different particles. The reason is that this constraint is concerted with
the light cone constraint \re{4.3} which, in turn, binds in an isotropic interval
the positions of even free particles. In Appendix B the dynamics of two free particles
is manifested from this tangled description.

The system determined by the set of 1st-class constraints \re{4.1},
\re{4.3} and \re{5.5} possesses 6 degrees of freedom.
Besides, as it follows from the structure of Lie algebra \re{5.1} of de Sitter group \cite{PWZ76},
of ten components of the conserved angular momentum tensor \re{4.14} one can construct
six integrals of motion which are in an involution in terms
of Poisson brackets. This is sufficient for the system to be
integrable in the Liouville sense. The next natural step would be a transition to
the description on a reduced 12-dimensional phase space, and
separating degrees of freedom by choosing appropriate canonical variables.
It turned out more constructive to analyze
the system within the manifestly covariant description on the
20-dimensional phase space ${\rm T}^*{\Bbb M}_5^2$ where de Sitter
symmetry is realized in a transparent way.


\section{Equations of motion and their integration.}
\renewcommand{\theequation}{6.\arabic{equation}}
\setcounter{equation}{0}

Useful integrals of motion arise from two Casimir functions of the de
Sitter algebra \re{5.1}:
%
\begin{equation} \label{6.2}
J^2:=-{\rm tr}({\s J}^2)=J_{MN}J^{MN},\qquad V^2:= V_MV^M,
\end{equation}
where the following 5-pseudo-vector
%
\begin{equation} \label{6.3}
V_M:=\frac18\epsilon_{MABCD}J^{AB}J^{CD}
\end{equation}
is introduced by means of the Levi-Chivita symbol
$\epsilon_{MABCD}$. Then using the equalities:
%
\begin{eqnarray}\label{7.1}
\Pi_{\!\bot}^2\approx-\frac1{R^2}\left(\ha
J^2+(\pi\!\cdot\!r)^2\right),\qquad
\pi_{\!\bot}^2\approx-\frac1{(\Pi\!\cdot\!r)^2R^2}\left(V^2-\ha(\pi\!\cdot\!r)^2J^2-(\pi\!\cdot\!r)^4\right)
\end{eqnarray}
recasts arguments of the dynamical constraint \re{5.5} into an
equivalent set,
%
\begin{equation} \label{7.2}
\Phi(\Pi\!\cdot\!r,\ \pi\!\cdot\!r;\ J^2,V^2)=0,
\end{equation}
which is more convenient for a dynamical analysis.

Let us consider the equation of motion for the relative position
5-vector $r$:
%
%
\begin{eqnarray}\label{eq1}
\dot r&=&\lambda\left\{r,\Phi(\Pi\!\cdot\!r,\,\pi\!\cdot\!r;\,
J^2,V^2)\right\}\nn\\
&=&\lambda\left(\frac{\partial\Phi}{\partial\,\pi\!\cdot\!r}
-4\frac{\partial\Phi}{\partial J^2}{\s
J}-\frac{\partial\Phi}{\partial V^2}{\s K}\right)r;
\end{eqnarray}
here ${\s J}:=||J^M_{\quad N}||$ and
%
\begin{equation} \label{6.3a}
{\s K}:=||K^M_{\quad N}||:=||\epsilon^M_{\ \ NABC}V^AJ^{BC}||
\end{equation}
are conserved matrices while the Lagrangian multiplier
$\lambda(\tau)$ as a function of $\tau$ is unspecified and can be
chosen for convenience reason.

If the variable $\Pi{\,\cdot\,}r=\Psi(\tau)$ was known as a function
of $\tau$ then $\pi{\,\cdot\,}r=\psi(\tau)$ could be found from the dynamical constraint \re{7.2} as
a solution of the algebraic equation:
$\Phi(\Psi(\tau),\psi(\tau);J^2,V^2)=0\quad\Longrightarrow\quad\psi(\tau):=\psi(\Psi(\tau);\
J^2,V^2)\quad$
(since $J^2$, $V^2$ are conserved).

In turn, the Hamiltonian equation for $\Psi(\tau)$,
%
\begin{eqnarray}\lab{7.8}
\dot
\Psi&=&\lambda\,\{\Psi,\Phi\}=\lambda\,\frac{\partial\Phi(\Psi,\
\psi(\Psi;J^2,V^2);\ J^2,V^2)}{\partial\psi}\,\Psi,
\end{eqnarray}
is self-sufficient, separable in $\tau$ and $\Psi$, and reduces obviously to quadratures.
Note that the resulting function $\Psi(\tau)$ depends on a choice of the Lagrange multiplier $\lambda(\tau)$.
Alternatively, one can choose $\Psi(\tau)$ and then find $\lambda(\tau)$ from eq. \re{7.8} without integration.
The choice of the function $\Psi(\tau)$ implies a fixing of the evolution parameter $\tau$.

    At this point the equation \re{eq1} becomes
a closed linear equation with respect to 5-vector $y$, with known
$\tau$-dependent matrix coefficients. The substitution
%
\begin{eqnarray}
r(\tau)&=&\frac{\Psi(\tau)}{\Psi(0)}\,q(\tau) \label{7.9}
\end{eqnarray}
simplifies this equation to the form:
%
\begin{eqnarray}
\dot q&=&-\lambda\left(4\frac{\partial\Phi}{\partial J^2}{\s
J}+\frac{\partial\Phi}{\partial V^2}{\s K}\right)q. \label{7.11}
\end{eqnarray}

A subsequent integration procedure is based on the projection
operator techniques described in Appendix C. A structure and the
action of projection operators depends on eigenvalues of the matrix
$\s J$ which, in turn, depend on values of the Casimir functions
\re{6.2}. Here we suppose $J^2<0$ and $V^2<0$ so that $\s J$ possesses
the following eigenvalues: $\pm\Sigma:=\pm\sqrt{\Sigma_+^2}$,
$\pm\im S:=\pm\sqrt{\Sigma_-^2}$ and 0, where $\Sigma_\pm^2$
are defined in \re{6.5}.
Other cases can be treated similarly; they are omitted here.

Let us decompose the 5-vector $q$ (and then other position
5-vectors) by means of the projection operators \re{6.6}-\re{6.9}
defined in Appendix C:
%
\begin{eqnarray}
q&=&({\cal O}^{(\Sigma)}+{\cal O}^{(S)}+{\cal P}^{(0)})q:=
q^{(\Sigma)}+q^{(S)}+q^{(0)}. \label{7.12}
\end{eqnarray}
The projectors \re{6.6}-\re{6.9} commute with the matrix ${\s J}$.
Using this fact and the properties \re{7.13}
of the matrix \re{6.3a} permits one to split the equation \re{7.11}
into the set:
%
\begin{eqnarray}
\dot q^{(i)}(\tau)&=&f^{(i)}(\tau){\s J}\,q^{(i)}(\tau),\qquad\qquad
i=\Sigma,S,0, \label{7.14}
\end{eqnarray}
where $f^{(0)}(\tau)\equiv0$,
%
\begin{eqnarray}
f^{(\Sigma)}(\tau):=-\lambda\left(4\frac{\partial\Phi}{\partial
J^2}+2S^2\frac{\partial\Phi}{\partial V^2}\right),\quad
f^{(S)}(\tau):=-\lambda\left(4\frac{\partial\Phi}{\partial
J^2}-2\Sigma^2\frac{\partial\Phi}{\partial V^2}\right).\qquad&&
\label{7.15}
\end{eqnarray}
%
Then formal solutions to the equations \re{7.14}-\re{7.15} are:
%
\begin{eqnarray}
q^{(i)}(\tau)&=&\exp\{F^{(i)}(\tau){\s J}\}r^{(i)}(0),
\quad\mbox{where}\quad
F^{(i)}(\tau):=\int_0^\tau\D{}{\tau}f^{(i)}(\tau), \quad
i=\Sigma,S,0.\qquad \label{7.16}
\end{eqnarray}
Matrix exponents in these solutions can be unraveled by means of
eqs. \re{6.10}:
%
\begin{eqnarray}
q^{(\Sigma)}(\tau)&=&\left(\cosh\left(\Sigma
F^{(\Sigma)}(\tau)\right)+\frac{\s J}{\Sigma}\sinh\left(\Sigma
F^{(\Sigma)}(\tau)\right)\right)r^{(\Sigma)}(0),
\label{7.17}\\
q^{(S)}(\tau)&=&\left(\cos\left(S F^{(S)}(\tau)\right)+\frac{\s J}{S}\sin\left(S F^{(S)}(\tau)\right)\right)r^{(S)}(0),\\
\label{7.18} q^{(0)}(\tau)&=&r^{(0)}(0). \label{7.18a}
\end{eqnarray}

A convolution of 5-vector $r$ with the angular momentum tensor \re{4.14}
expressed in terms of the collective variables \re{5.0}
yields the equality for the
5-vector $Y$:
%
\begin{eqnarray}\label{7.19}
Y&\approx&\frac{{\s J}-\psi}{\Psi}\,r.
\end{eqnarray}
Then eqs. \re{7.9}, \re{7.12}, \re{7.17}--\re{7.19} lead to the
expressions for particle positions
%
\begin{eqnarray}
y_a^{(i)}(\tau)&=&\frac1{\Psi(0)}\left\{{\s
J}-\psi(\tau)-\ha(-)^a\Psi(\tau)\right\}q^{(i)}(\tau),\qquad
a=1,2,\quad i=\Sigma,S,0,\quad \label{7.20}
\end{eqnarray}
where all the quantities in r.-h.s. are known functions of $\tau$ at
this point.

In order to have a complete solution for Cauchy problem, it is
sufficient to express the angular momentum matrix ${\s J}$ and its
invariants $\Sigma$, $S$ in terms of initial values $y_a(0)$, $\dot
y_a(0)$ by eqs. \re{4.14}, \re{4.15} and \re{6.2}, \re{6.3}, \re{6.5}.
If the initial point belongs to
${\rm T}\,{\Bbb K}$, i.e., the  initial values $y_a(0)$, $\dot y_a(0)$
are subjected to the conserved holonomic constraints \re{4.1}, \re{4.3}
and their differential consequences (see also \re{A.1}, \re{A.2}), then the
particle world lines \re{7.20} lie in ${\Bbb K}$ by construction.  The momentum-type
variables \re{5.3}, \re{5.4} are subsidiary and not important within
the classical consideration.


\section{Conclusion}

Green functions of massless fields in the Minkowski space-time are
located on the light cone surface. This field-theoretical outcome
was basic for a construction of the original
Staruszkiewicz-Rudd-Hill model and its non-electromagnetic
generalizations.

In the curved space-time the Green function of electromagnetic and
other massless fields possesses a non-local tail spread in the light
cone interior \cite{PPV11}. It is shown here that in particular
case of de Sitter space-time the nonlocal contribution of the
electromagnetic Green function in the Tetrode-Fokker action integral
can be converted to a dynamically equivalent local contribution. The
nonlocal contribution of the scalar Green function is unavoidable,
if the theory of minimal coupling is implied. Instead, the Green
function of the scalar field conformally coupled to de Sitter
metrics is shown to be purely local. These two examples of
field-theoretical nature are included in a wide class of
time-asymmetric models built from general demands of
de Sitter symmetry and self-consistency of the Hamiltonian dynamics.

Every time-asymmetric model possesses 6 degrees of freedom and
6 integrals of motion in involution which are independent
functions of canonical generators $J_{MN}$ of O(1,4) group \cite{PWZ76}.
Thus these dynamical systems are integrable in the Liouville sense.
In practice, the integrability presupposes a choice of appropriate canonical
variables in terms of which degrees of freedom separates.
In the case of curved de Sitter space-time this task encounters technical difficulties
when constructing the description in a 12-dimensional phase space.

Thus in the present paper the time-asymmetric models are treated as
constrained systems in 20-dimensional phase space $\mathrm{T}^*{\Bbb
M}^2_5$. de Sitter invariance of all the constraints admits a
formulation of equations of motion in a manifestly covariant
5-dimensional form. Moreover, there exists some analogy between a
dynamics of the relativistic particle in a constant electromagnetic
field \cite{Fra78,Yar13} and the present problem. As the Maxwell tensor
in the first case, the conserved angular 5-momentum tensor in the second case
is skew-symmetric, treated as constant and covariantly ``mounted'' into equations of motion.
Thus the projection operator technique, used in the first case
\cite{Fra78,Yar13}, is adapted here to the present 5-dimensional
case. In such a way the equations of motion are split and solved in
quadratures.

It was noted in Introduction that the Staruszkiewicz-Rudd-Hill model
in a flat space-time endow corresponding two-particle systems with
physically meaningful features. What distinguishes the model from
the retarded or Wheeler-Feynman electrodynamics is
the time-asymmetric retarded-advanced causal structure of
interaction, a price for a solvability of the model.
Even so, the classical model represents properly relativistic effects
in a system of two charged particles within the moderately relativistic domain
where the radiation reaction is minor.
The quantum versions of this model and some other
time-asymmetric models yield relativistic spectra which accord well
with results of the quantum field theory \cite{D-S01} and actual
meson spectroscopy \cite{Duv01}.

A study of de-Sitter-relativistic effects in systems of single gravitating bodies and
test particles \cite{AAMP07,CGKM08,Cas08} deepen understanding of
the expanding Universe. The next step in this direction would be
a prospective elaboration of de Sitter invariant two-particle models with
electromagnetic and other interactions. A quantization of time-asymmetric models
in de Sitter space is addressed to future works.

\section*{Acknowledgments}

The author thanks to Yu. Yaremko for fruitful discussions.


\section*{Appendix}

\subsection*{A. The relation between the Lagrangian function and the dynamical constraint
of time-asymmetric models}
\renewcommand{\theequation}{A.\arabic{equation}}
\setcounter{equation}{0}

Chosen the sign $\eta=1$ or $\eta=-1$  in the model (see Section 4), let us present the Lagrangian \re{4.5} in the
equivalent form:
%
\begin{eqnarray}
&&L = \vartheta F(\nu_1, \nu_2, \omega), \label{4.11}\\
&&\vartheta := \eta \dot y_1 \!\cdot r =
              \eta \dot y_2 \!\cdot r = \eta(\dot y_1+\dot y_2)\!\,\cdot r/2 > 0,
\label{4.13}\\
&&F := -\sum_{a=1}^2 \frac{\eta m_a}{R\nu_a} - \frac{f(\nu_1,\nu_2,\omega)}{R^2\nu_1\nu_2}, \label{4.12}
\end{eqnarray}
where $F$ is a function of the scalar arguments \re{3.8} and thus is a homogeneous function of degree zero of particle velocities $\dot y_a$.
The scalar factor $\vartheta$ is homogeneous of degree one and positive on timelike world lines.
It is presented in \re{4.13} diversely
by accounting a differential consequence  $\dot y_1\cdot r=\dot y_2\cdot r$ of the
light cone constraint \re{4.3}.
In this regards an apparent particle asymmetry of the interaction term of the Lagrangian \re{4.5} is seeming.

In terms of the functions \re{4.13}, \re{4.12} and the collective variables \re{5.0}
the Legendre transform \re{4.15} acquires the manifestly covariant 5-vector form:
%
\begin{eqnarray}
\Pi &=& \frac{\partial L}{\partial\dot Y}
= A(\nu_1,\nu_2,\omega)\frac{\dot Y}{\vartheta} + B(\nu_1,\nu_2,\omega)\frac{\dot r}{\vartheta}
+D(\nu_1,\nu_2,\omega)\eta r,
\label{5.6a}\\
\pi &=& \frac{\partial L}{\partial\dot y}
= B(\nu_1,\nu_2,\omega)\frac{\dot Y}{\vartheta} + C(\nu_1,\nu_2,\omega)\frac{\dot r}{\vartheta},
\label{5.6b}
\end{eqnarray}
where
\begin{eqnarray}
A&:=&A_1+A_2,\quad B:=\frac{A_1-A_2}2,\quad C:= \frac{A}4+\frac{\partial f}{\partial\omega},\quad
D :=\frac1{R^2}\left[\frac{f}{\nu_1\nu_2}-\frac1\nu_2\frac{\partial f}{\partial\nu_1}-\frac1\nu_1\frac{\partial f}{\partial\nu_2}\right],
\nn\\
A_a&:=& -\eta R m_a\nu_a-\frac{\nu_a}{\nu_{\bar a}}f
+\frac{\nu_a^2}{\nu_{\bar a}}\frac{\partial f}{\partial\nu_a}
+\left[\frac{\nu_a}{\nu_{\bar a}}\omega-1\right]\frac{\partial f}{\partial\omega};\qquad {a=1,2,\quad\atop\bar a=3-a.}\hspace{4ex}
\label{5.7b}
\end{eqnarray}
The right-hand side of eqs. \re{5.6a}, \re{5.6b} is evidently zero-degree homogeneous in $\dot Y$, $\dot y$, thus the Legendre transform is degenerated.

We are interested in relations between scalars on ${\mathrm T}^*{\Bbb M}_5^2$ and ${\mathrm T}{\Bbb K}$,
where ${\mathrm T}{\Bbb K}\subset{\mathrm T}{\Bbb M}_5^2$ is described by the holonomic constraints \re{4.1}, \re{4.3}
and their differential consequences expressed for a conveniency in terms the collective variables \re{5.0}:
%
\begin{align}
&Y^2=-R^2; &&Y\cdot r=0;  &&r^2=0,\ \eta r^0>0;
\label{A.1}\\
&\dot Y\cdot Y=0; &&\dot Y\cdot r=-\dot r\cdot Y; &&\dot r\cdot r=0.
\label{A.2}
\end{align}
Multiplying eqs. \re{5.6a},  \re{5.6b} by $r$ and $Y$ and accounting \re{A.1}, \re{A.2} yields the relations:
%
\begin{align}
\Pi\cdot r&=\eta A(\nu_1,\nu_2,\omega); & \pi\cdot r&=\eta B(\nu_1,\nu_2,\omega);
\label{A.3}\\
\Pi\cdot Y&=-\eta B(\nu_1,\nu_2,\omega); & \pi\cdot Y&=-\eta C(\nu_1,\nu_2,\omega).
\label{A.4}
\end{align}
Among these scalars on ${\mathrm T}^*{\Bbb M}_5^2$ two of them, $\Pi\cdot r$ and $\pi\cdot r$, are
observables, and they are arguments of the dynamical constraint \re{5.5}.
Squaring eq. \re{5.6a}, one can express the scalar $\Pi^2$ in terms of $\nu_1$, $\nu_2$, $\omega$.
Scalars $\Pi\cdot Y$, $\pi\cdot Y$ and $\Pi^2$ are not
the observables, but they are related with the third argument $\Pi_\bot^2$ of \re{5.5} via the following
equality derived by squaring eq. \re{5.3}:
$
\Pi_\bot^2=\Pi^2+ [(\Pi\cdot Y)^2+2(\Pi\cdot r)(\pi\cdot Y)]/R^2.
$
Using this and previous equations yields:
%
\begin{eqnarray}
\Pi_\bot^2=\frac1{R^2}\left[\frac{A^2_1}{\nu^2_1}+\frac{A^2_2}{\nu^2_2}+2\omega\frac{A_1A_2}{\nu_1\nu_2} + B^2-2AC\right]+ 2AD.
\label{A.5}
\end{eqnarray}

In general, three equations \re{A.3} and \re{A.5} can be inverted yielding
$\nu_1$, $\nu_2$ and $\omega$ as functions of $\Pi\cdot r$, $\pi\cdot r$ and $\Pi_\bot^2$.
We will use for these functions the notations $\bar\nu_1$, $\bar\nu_2$, $\bar\omega$,
and $\bar A:=A(\bar\nu_1,\bar\nu_2,\bar\omega)=\eta\Pi\cdot r$, \dots, $\bar D:=D(\bar\nu_1,\bar\nu_2,\bar\omega)$
etc. At this point the set of equations \re{5.6a}, \re{5.6b} can be formally inverted
yielding velocities in terms of canonical variables:
%
\begin{eqnarray}
\frac{\dot Y}\vartheta &=& \frac{\bar C}{\bar\Delta}(\Pi-\bar D\eta r) - \frac{\bar B}{\bar\Delta}\pi,\qquad
\frac{\dot r}\vartheta = \frac{\bar A}{\bar\Delta}\pi - \frac{\bar B}{\bar\Delta}(\Pi-\bar D\eta r),
\label{A.6}
\end{eqnarray}
where $\Delta:=AC-B^2$. Then the l.h.-s. of eq. \re{4.16} can be regarded as the Hamiltonian, proportional to the dynamical constraint: $H_{\rm D}\propto\Phi=0$.
Inserting the expressions \re{A.6} for the particle velocities $\dot y_a=\dot Y-\ha(-)^a\dot r$ ($a=1,2$)
into l.h.-s. of eq. \re{4.16} yields the dynamical constraint of the form \re{5.5}:
%
\begin{eqnarray}
\pi_\bot^2+\frac{\bar C^2}{R^2}+\frac{\bar C}{\eta\Pi\cdot r}\left[\Pi_\bot^2-\frac{(\pi\cdot r)^2}{R^2}\right]
- \frac{\bar\Delta(\bar F+\bar D)}{\eta\Pi\cdot r}=0.
\label{A.7}
\end{eqnarray}
It determines the scalar observable $\pi_\bot^2$ via three other arguments $\Pi\cdot r$, $\pi\cdot r$ and $\Pi_\bot^2$
of the dynamical constraint \re{5.5}. In the free-particle case $f=0$ we arrive at eq. \re{C.1a}.

One can obtain other relations between canonical variables,
such as $\Pi\cdot Y+\pi\cdot r=0$, following from \re{A.3}, \re{A.4}.
These relations represent secondary constraints, mentioned in Section 5, which
involve unobservable quantities and thus have not a physical meaning.


\subsection*{B. The free-particle system}
\renewcommand{\theequation}{B.\arabic{equation}}
\setcounter{equation}{0}

The free-particle dynamical constraint \re{C.1a} can be presented diversely:
%
\begin{eqnarray}
\Phi_{\rm free}
&\approx& \frac{\pi_2\!\cdot\!r}{\Pi\!\cdot\!
r}\phi_1+\frac{\pi_1\!\cdot\!r}{\Pi\!\cdot\! r}\phi_2=0,
\label{C.1}
\end{eqnarray}
where $\phi_a:={\pi_a^2}_{\!\bot}-m_a^2$ and
$\displaystyle{{\pi_a}_{\!\bot}:= \pi_a
-\frac{y_a\cdot\pi_a}{y_a^2}y_a}$ ($a=1,2$).
This form is more convenient here. It does not imply, however, that
both the expressions $\phi_1$ and $\phi_2$ vanish (as one could
opine by Section 2), so that
$\phi_-:=\ha(\phi_1-\phi_2)\ne0$.

The Hamilton equations for the position 5-vectors read:
%
\begin{equation} \label{C.2}
\dot y_a=\lambda\{y_a,\Phi_{\rm
free}\}\approx2\lambda\frac{\pi_{\bar a}\!\cdot\!r}{\Pi\!\cdot\! r}
\left({\pi_a}_{\!\bot}+(-)^a\frac{\phi_-}{\Pi\!\cdot\!
r}r\right),\qquad {a=1,2,\quad\atop\bar a=3-a,}
\end{equation}
and yield the expressions for the  unit 5-velocities of particles:
%
\begin{equation} \label{C.4}
v_a\equiv\frac{\dot y_a}{\sqrt{\dot
y_a^2}}\approxeq\frac1{m_a}\left({\pi_a}_{\!\bot}+(-)^a\frac{\phi_-}{\Pi\!\cdot\!
r}r\right)
\end{equation}
which are free of the unspecified Lagrangian multiplier $\lambda$;
here the symbol ``~$\approxeq$~'' denotes a weak equality by virtue
of all the constraints  \re{4.1}, \re{4.3} and \re{C.1a}.

Differentiating the equalities \re{C.4} and using the Hamiltonian
equations \re{C.2} and corresponding  equations for $\pi_a$ yields
the expressions for derivatives $\dot v_a$:
%
\begin{equation} \label{C.5}
\dot v_a\approxeq2\lambda\frac{\pi_{\bar a}\!\cdot\!r}{\Pi\!\cdot\!
r}\frac{\sqrt{\dot y_a^2}}{m_a}{R^2}y_a.
\end{equation}
From \re{C.2} and \re{C.5} the 2nd-order equations of motion follow:
%
\begin{equation}\lab{C.6}
\frac{\D{}{ }}{\D{}{\tau}}\frac{\dot y_a}{\sqrt{\dot
y_a^2}}-\sqrt{\dot y_a^2}\frac{y_a}{R^2}=0, \qquad a=1,2.
\end{equation}
They are split in variables of different particles, and coincide for
each particle with the test body equation \re{2.5}. The solutions $y_a(\tau)$ have the form
\re{2.6}, \re{2.6a} for each particle $a=1,2$.


\subsection*{C. The angular 5-momentum tensor and projection operators}
\renewcommand{\theequation}{C.\arabic{equation}}
\setcounter{equation}{0}

Components of the angular 5-momentum tensor form the skew-symmetric
odd-dimensional matrix $||J_{MN}||$, thus one of its
eigenvalue is zero. The same is concerned with the matrix $\s
J:=||J^M_{\ \ N}||:=||\eta^{ML}J_{LN}||$. In order to find other eigenvalues of
$\s J$ one can use the Hamilton-Cayley theorem and construct for
$\s J$ the characteristic equation. It obviously includes odd degrees of
$\s J$ up to five with de Sitter invariant coefficients.
Then one arrives by direct calculations at the desirable identity:
%
\begin{equation} \label{6.4}
{\s J}^5+\ha J^2{\s J}^3+V^2{\s J}\equiv0,
\end{equation}
where $J^2$ and $V^2$ are two Casimir functions of de Sitter
algebra, defined by eqs. \re{6.2}, \re{6.3}. The l.-h.s. of \re{6.4}
can be formally factorized:
%
\begin{equation} \label{6.1}
({\s J}^2-\Sigma^2_+)({\s J}^2-\Sigma^2_-){\s J}=0,
\end{equation}\vspace{-6ex}

\noindent
where
%
\begin{equation} \label{6.5}
\Sigma^2_\pm:=-\qu J^2\pm\sqrt{\cal D},\qquad
{\cal D}:=J^4/{16}-V^2.
\end{equation}
Thus the matrix $\s J$ possesses 5 eigenvalues $\pm\Sigma_+,
\pm\Sigma_-,0$.

Projection operators onto 1-dimensional subspaces corresponding to
eigenvalues $j$ of $\s J$ can be introduced
by a standard technique; see for example \cite{Yar13}:
%
\begin{equation} \label{B.1}
{\cal P}^{(j)}=\prod\limits_{j^\prime\ne j}\frac{{\s J}-j^\prime}{j-j^\prime},
\qquad j=\pm\Sigma_+,\pm\Sigma_-,0;
\end{equation}
here $j^\prime$ in the product runs over all eigenvalues except $j$.

In general, the Casimir functions $J^2$ and $V^2$ and thus the discriminant
${\cal D}$ can acquire arbitrary real (positive or negative) values, so that
the eigenvalues $j$ can be real or complex. Here, however, we limit this
arbitrariness by natural physical restrictions.

For the single-particle case $J^2\approx-m^2R^2$ while $V=0$. In the
case of two free particles one obtains from \re{4.14}, \re{4.15} and  \re{4.5} (with $f=0$):
%
\begin{eqnarray}
\varkappa&:=&-\frac{J^2}{4m_1m_2R^2}=\mu+\omega-\nu_1\nu_2,
\label{B.2}\\
\chi&:=&\frac{V^2}{m_1^2m_2^2R^4}=\nu_1^2+\nu_2^2-2\nu_1\nu_2\omega+\nu_1^2\nu_2^2,
\label{B.3}\\
\delta&:=&\frac{\cal D}{m_1^2m_2^2R^4}=\varkappa^2-\chi=(\mu+\omega)^2-\nu_1^2-\nu_2^2-2\mu\nu_1\nu_2,
\label{B.4}
\end{eqnarray}
where $\omega$ and $\nu_a$ ($a=1,2$) are defined by eq. \re{3.8}, and $\mu:=\frac12[\frac{m_1}{m_2}+\frac{m_2}{m_1}]\ge1$.

Let us evaluate $J^2$ and $\cal D$ (or $\varkappa$ and $\delta$) on the time-like world lines, for which
$v_a^2=1$, $v_a^0\ge1$ ($a=1,2$).
Since the Casimir functions are integrals of motion and O(1,4)-invariants, it is sufficient to evaluate r.h.-s. of
\re{B.2}-\re{B.4} at the initial moment $\tau=0$ in arbitrary reference frame.

We will use for 5-vectors the 3-vector notations: $y=\{y^0,y^1,y^2,y^3,y^4\}:=\{y^0,\B y,y^4\}$.

Let us start with the case $\eta=+1$, i.e., $y_1^0>y_2^0$.

The action of the group O(1,4) on the hyperboloid $\Bbb H$ is transitive \cite{CGKM08}.
Thus there exists a reference frame where the starting 5-position $y_1$ of the 1st particle and its 5-velocity $v_1$
are as follows:
%
\begin{equation} \label{B.5}
y_1=\{0,\B0,R\}, \qquad v_1=\{1,\B0,0\}.
\end{equation}
Thus $\omega=v_2^0\ge1$. Besides, it follows from \re{B.5} and the constrains  \re{4.1}, \re{4.3}: $y_2^4=R$, $y_2^0=-|\B y_2|$
with arbitrarily chosen 3-vector $\B y_2$, i.e.,
%
\begin{equation} \label{B.6}
y_2=\{-|\B y_2|,\B y_2,R\}.
\end{equation}
Now, using the differential consequence $y_2\cdot v_2=0$ of the constrains \re{4.1} yields $v_2^4<0$.
Thus $\nu_1=-y_2\cdot v_1=|\B y_2|/R>0$, $\nu_2=y_1\cdot v_2=-v_2^4>0$, and $\omega^2-\nu_2^2\ge1$, $\omega-\nu_2>0$.

Finally we impose the additional condition $y^0+y^4>0$. It selects a half of the hyperboloid $\Bbb H$ which
corresponds to the flat exponentially expanding Friedmann universe \cite{CGKM08}. It is obviously from \re{B.5}
$y_1^0+y_1^4>0$. If the second particle belongs to the same universe, i.e., $y_2^0+y_2^4>0$, then the restriction $|\B y_2|<R$
follows from \re{B.6}. Thus $\nu_1<1$. Using all these inequalities yields the estimates:
\begin{eqnarray*}
\varkappa&=&\mu+\omega-\nu_1\nu_2>\mu+\omega-\nu_2>\mu,
\\
\delta&=&(\mu+\omega)^2-\nu_1^2-\nu_2^2-2\mu\nu_1\nu_2>(\mu+\omega)^2-\omega^2-2\mu\omega=\mu^2,
\end{eqnarray*}
%
so that
%
\begin{eqnarray}
J^2<-2(m_1^2+m_2^2)R^2<0,\quad {\cal D}>(m_1^2+m_2^2)^2R^4/4>0\ \Rightarrow\
\Sigma_+^2>(m_1^2+m_2^2)R^2>0\qquad\quad
\label{B.7}
\end{eqnarray}
while $\Sigma_-^2$ can be negative or positive.

For $\eta=-1$ the same estimates can be obtained by
the particle permutation $1\leftrightarrow2$.

If an interaction of particles is present but not too strong to close up the gaps $\propto m_1^2+m_2^2$ in \re{B.7},
the inequalities $J^2<0$, ${\cal D}>0$ may hold, and we have again $\Sigma_+^2>0$ and $\Sigma_-^2\lessgtr0$.

Here we consider the case $\Sigma_+^2:=\Sigma^2>0$, $\Sigma_-^2:=-S^2<0$ in detail.
The matrix $\s J$ possesses 5 eigenvalues: $\pm\Sigma$, $\pm\im
S$ (where $\Sigma>S>0$) and 0.

Projection operators \re{B.1} onto 1-dimensional subspaces corresponding to
these eigenvalues have the form:
%
\begin{eqnarray}
{\cal P}^{(\pm\Sigma)}&:=&\frac{({\s J}\pm\Sigma)({\s J}^2+S^2){\s
J}}{2\Sigma^2(\Sigma^2+S^2)},\quad {\cal P}^{(\pm\im S)}:=\frac{({\s
J}\pm\im S)({\s J}^2-\Sigma^2){\s J}}{2S^2(\Sigma^2+S^2)},
\label{6.6a}\\
{\cal P}^{(0)}&:=&-\frac{({\s J}^2+S^2)({\s
J}^2-\Sigma^2)}{\Sigma^2S^2}.
\label{6.6}
\end{eqnarray}
Instead of projectors \re{6.6a}, it is convenient to use analogs of
Fradkin operators \cite{Fra78,Yar13}:
%
\begin{eqnarray}
{\cal O}^{(\Sigma)}&:=&{\cal P}^{(+\Sigma)}+{\cal
P}^{(-\Sigma)}=\frac{({\s J}^2+S^2){\s
J}^2}{\Sigma^2(\Sigma^2+S^2)},
\nn\\
{\cal O}^{(S)}&:=&{\cal P}^{(+\im S)}+{\cal P}^{(-\im S)}=\frac{({\s
J}^2-\Sigma^2){\s J}^2}{S^2(\Sigma^2+S^2)} \label{6.9}
\end{eqnarray}
which project onto the corresponding 2-dimensional subspaces. We note
the important properties of these operators:
%
\begin{equation} \label{6.10}
{\s J}^2{\cal O}^{(\Sigma)}=\Sigma^2{\cal O}^{(\Sigma)}, \qquad {\s
J}^2{\cal O}^{(S)}=-S^2{\cal O}^{(S)}, \qquad {\s J}\,{\cal
P}^{(0)}=0.
\end{equation}

In order to derive important properties of the matrix $\s K$ defined by eq. \re{6.3a} it should be
simplified. Accounting \re{6.3} in \re{6.3a} and unraveling the convolution of Levi-Civita symbols
$\epsilon_{..\dots}\epsilon^{..\dots}$ in terms of
products of Kronecker symbols $\delta_.^.\cdots\delta_.^.$ yields the formula:
%
\begin{equation} \label{6.10a}
{\s K}=2{\s J}^3+J^2 {\s J}.
\end{equation}
The action of the projectors \re{6.6}, \re{6.9} onto \re{6.10a} results in the relations:
%
\begin{eqnarray} \label{7.13}
{\cal O}^{(\Sigma)}{\s K}=2S^2{\cal O}^{(\Sigma)}{\s J}, \qquad
{\cal O}^{(S)}{\s K}=-2\Sigma^2{\cal O}^{(S)}{\s J},\qquad {\cal
P}^{(0)}{\s K}=0.
\end{eqnarray}
The properties \re{6.10} and \re{7.13} are used in Section 7 for the integration of the system.

The case $\Sigma_+^2>0$, $\Sigma_-^2>0$ can be considered similarly.


\end{document}